\documentclass[11pt]{article}
\usepackage{amsfonts}
\usepackage{amsmath}
\usepackage{amssymb}
\usepackage{epsfig}
\usepackage{graphicx}
\usepackage{mathptmx}
\usepackage{multirow}
\usepackage[colorlinks=true,linkcolor=red,citecolor=blue]{hyperref}
\usepackage{url}
\begin{document}
\title{Charmless  $B_{(s)}\to VV$ Decays in Factorization-Assisted Topological-Amplitude Approach}
\author{ Chao Wang$^{1}$, Qi-An Zhang$^{1}$, Ying Li$^{2,3}$\footnote{Email:liying@ytu.edu.cn},
Cai-Dian L\"u$^{1,3}$\footnote{Email:lucd@ihep.ac.cn}\\
{\small \it 1.~Institute of High Energy Physics, CAS, P.O. Box 918, Beijing 100049, China  }\\
{\small \it and School of Physics, University of Chinese Academy of Sciences, Beijing 100049, China} \\
{\small \it 2.~Department of Physics, Yantai University, Yantai 264005, China}\\
{\small \it 3.~State Key Laboratory of Theoretical Physics, Institute of Theoretical Physics,CAS, Beijing 100190, China}
}
\maketitle
\begin{abstract}
Within the factorization-assisted topological-amplitude approach, we studied the 33 charmless $B_{(s)} \to VV$   decays, where $V$ stands for a light vector meson. According to the flavor flows, the amplitude of each process can be decomposed into 8 different topologies. In contrast to the conventional flavor diagrammatic approach, we further factorize each topological amplitude into decay constant, form factors and unknown universal parameters. By  $\chi^2$ fitting  46 experimental observables, we extracted   10 theoretical  parameters with $\chi^2$ per degree of freedom around 2. Using the fitted parameters, we calculated the branching fractions, polarization fractions, CP asymmetries and relative phases between polarization amplitudes of each decay mode.  The decay channels dominated by tree diagram have large branching fractions and large longitudinal polarization fraction. The branching fractions and longitudinal polarization fractions  of color-suppressed decays become smaller. 
Current experimental data of large transverse polarization fractions in the penguin dominant decay channels can be explained by only one transverse amplitude of penguin annihilation diagram. Our predictions of those not yet measured channels can be tested in the ongoing LHCb experiment and the Belle-II experiment in future.
\end{abstract}

\section{Introduction}\label{sec-1}

Charmless hadronic $B$-meson decays have been of significant interest, as they can provide us an abundant source of information on the flavor physics, within and beyond the standard model (SM). After the very successful first generation of $B$-factory experiments, BaBar and Belle \cite{Bevan:2014iga}, the interest in this field is reinforced by the LHCb experiment \cite{Buchalla:2008jp} and the upcoming start of Belle-II experiment \cite{Aushev:2010bq,Browder:2007gg}. In the long term run, the LHCb upgrade plan promise excellent future opportunity\cite{Gianotti:2002xx,LHCb:2011dta}. The FCC-ee, as well as CEPC, the proposal for future electron-positron collider will give further chance for the flavor physics study \cite{cepc}.
In the theoretical side, beyond the naive factorization approach \cite{Wirbel:1985ji}, three major QCD-inspired approached had been proposed to deal with charmless nonleptonic $B$ decays, based on the effective theories,  namely,  the QCD factorization (QCDF) \cite{Beneke:2000ry}, perturbative QCD (PQCD) \cite{Lu:2000em}, and soft-collinear effective theory (SCET) \cite{Bauer:2000yr}.   The difference between them  is only on the treatment of dynamical degrees of freedom at different mass scales, namely the power counting. Within these approaches, most decay processes have been studied, including the branching fractions and the $CP$ asymmetries. However,
the factorization for hadronic matrix elements   is only proved at the leading order in $1/m_b$, with   $m_b$ denoting     the $b$ quark mass.
The precision of these approaches is limited to the leading power calculations.

In contrast to the above approaches based on the perturbative QCD, another idea  based on the topological diagrams and flavor  SU(3) symmetry   was also proposed \cite{diagramapp},   where  the nonperturbative parameters are   extracted directly from experimental data. Therefore, the extracted parameters include the effects of strong interactions to all orders, as well as long-distance rescattering. This idea has been used to analyze hadronic $B$ meson decays extensively \cite{cwchiangB}, as well as $D$ meson decays \cite{Cheng:2012xb}. Although no direct power expansion is needed in this approach, flavor SU(3) symmetry is required to reduce the number of free parameters to be fitted from experiments. As the experimental precision is better and better, the limitation of theoretical precision is retarded.  Recently, the improved version, the so-called factorization-assisted topological-amplitude  approach (FAT) \cite{Li:2012cfa} was proposed, in order to deal with the SU(3) breaking effects. By using some of  the well defined factorization  formulas to include  most of the SU(3) breaking effects, the theoretical results of the two body non-leptonic $D$ decays accommodate experimental data very well. Recently, the FAT approach has been utilized to study the two-body charmed nonleptonic $B$ mesons decays    \cite{Zhou:2015jba}. Within 4 universal non-perturbative parameters fitted from 31 experimental observations, 120 charmed $B$ decay modes were calculated.   Both branching fractions and $CP$ asymmetry parameters are in agreement with experimental data well. Very recently, the charmless $B_{(s)} \to PP$ and $B_{(s)}\to PV$ processes are  also studied using this approach  \cite{Zhou:2016jkv}. The long-standing $B\to \pi^0 \pi^0$ and $B\to K\pi$ $CP$ puzzles can be explained simultaneously.

 In contrast to $B_{(s)} \to PP$ and $PV$ decays, charmless $B_{(s)} \to V V$ decays are much more complicated, because more helicity amplitudes will be considered. Due to angular momentum conservation, there are three independent configurations of the final-state spin vectors: a longitudinal component where   both resonances are polarized in their direction of motion, and two transverse components with perpendicular and transverse polarizations. For the $V-A$ coupling of the SM, a specific pattern of the three helicity amplitudes is naively expected~\cite{Koerner1979}, such that the longitudinal polarization fraction $f_L$ should be close to unity, while the transversal contributions are suppressed by $\Lambda_{QCD}/m_B$. In 2004,   large transverse polarization fractions (around 50\%) of  $B\to \phi K^*$  have been measured in the experiments. Later on, some other penguin-dominated strangeness-changing decays, such as $B\to \rho K^*$ and $B_s\to \phi \phi$, have also been found with large transverse polarization fractions. These large unexpected transverse polarization fractions have attracted much theoretical attention with several explanations based on the QCDF \cite{Li:2003he, Kagan:2004uw, Beneke:2006hg, Bartsch:2008ps, Cheng:2008gxa, Cheng:2009cn, Cheng:2009mu}, PQCD \cite{Li:2004ti, Ali:2007ff, Zou:2015iwa}, even on the new physics scenarios \cite{Hou:2004vj, Ladisa:2004bp, Yang:2004pm, Das:2004hq, Kim:2004wq, Zou:2005gw, Huang:2005if, Baek:2005jk, Huang:2005qb, Bao:2008hd}. These decays have     rich observables, some of which are regarded as good places for testing the SM and searching for possible effects of new physics beyond the SM.

To our knowledge, these decays with two vector meson final states have not been studied in the flavor diagram approach.
In this work, we shall  explore the charmless two-body non-leptonic $B_{(s)} \to V V$  decays, in the newly established  FAT  approach.   The branching fractions, CP asymmetries, as well as the angular distributions will be investigated. The organization of the paper is as follows: In Section~\ref{sec-2}, we   give the definitions for   helicity amplitudes, angular variables and polarization observables. The calculation of the $B\to VV$ decay amplitudes in FAT framework is briefly reviewed. Section~\ref{sec-3} provides the  numerical results and the phenomenological discussions. We will summarize work in Section~\ref{sec-4}.

\section{Framework}\label{sec-2}


We consider a $B$ meson with four-momentum $p_B$ decaying into two vector mesons $ V_1(m_1, p_1, \eta^*)$, $ V_2(m_2,p_2,\epsilon^*)$, where $\eta^*$ and $\epsilon^*$ are the polarization vectors of each final state meson. The decay amplitude can thus be decomposed into three parts,
\begin{eqnarray}
{\cal A}_{B\to V_1V_2}=i\eta^{*\mu} \epsilon^{*\nu}\left(g_{\mu\nu}S_1-\frac{p_{B\mu}p_{B\nu}}{m_B^2}S_2+i\varepsilon_{\mu\nu\rho\sigma}\frac{p_1^\rho p_2^\sigma}{p_1\cdot p_2}S_3\right).
\end{eqnarray}
With definite helicity, we can write the amplitudes as :
\begin{eqnarray}
&&{\cal A}^0={\cal A}(B\to V_1( p_1, \eta_0^*)V_2( p_2, \epsilon^*_0))=i\frac{m_B^2}{2m_1m_2}\left(S_1-\frac{S_2}{2}\right),\\
&&{\cal A}^\pm={\cal A}(B\to V_1( p_1, \eta_\pm^*)V_2( p_2, \epsilon^*_\pm))=i\left(S_1\mp S_3\right).
\end{eqnarray}

In the naive factorization, the helictiy amplitudes ${\cal A}^h_{ B\to V_1V_2}$ are proportional to
\begin{eqnarray}
 {\cal A}^h=\frac{G_F}{\sqrt 2}V_{CKM}\langle V_1^h|(\bar{b} q_s)_{V-A}| B\rangle\langle V_2^h|(\bar q q^\prime)_V|0\rangle .
\end{eqnarray}
If we ignore the  $m_i^2$ terms, the above functions can be simplified as
\begin{eqnarray}
{\cal A}^0&=&i\frac{G_F}{\sqrt 2}V_{CKM}f_2m_B^2A_0^{BV_1}(0),\\
{\cal A}^+&=&i\frac{G_F}{\sqrt 2}V_{CKM}f_2m_2\left\{- \left(m_B+m_1\right) A_1^{BV_1}(0)+ \left(m_B-m_1\right)V^{BV_1}(0)\right\}, \\
{\cal A}^-&=&i\frac{G_F}{\sqrt 2}V_{CKM}f_2m_2\left\{ -\left(m_B+m_1\right) A_1^{BV_1}(0)- \left(m_B-m_1\right)V^{BV_1}(0)\right\},
\end{eqnarray}
where $A_0^{BV_1}(0)$, $A_1^{BV_1}(0)$ and $V^{BV_1}(0)$ are transition form factors, the definitions of which   are given in ref.\cite{Wirbel:1985ji}. It is apparent that the transverse amplitudes ${\cal A}^\pm$ are suppressed by a factor $m_2/m_B$ relative to  ${\cal A}^0$. Due to the cancelation between the axial vector form factor $A_1$ and the vector  form factor $V$, we can arrive the relations as
\begin{eqnarray} \label{herachy}
{\cal A}^0:{\cal A}^-:{\cal A}^+=1:\frac{\Lambda_{QCD}}{m_b}:\left(\frac{\Lambda_{QCD}}{m_b}\right)^2.
\end{eqnarray}
Alternatively, we  can also adopt the transversity convention to get:
\begin{eqnarray}
{\cal A}_{L}={\cal A}^0;\,\,\,\,\,
{\cal A}_{\parallel}=\frac{{\cal A}^++{\cal A}^-}{\sqrt 2};\,\,\,\,\,
{\cal A}_{\perp}=\frac{{\cal A}^+-{\cal A}^-}{\sqrt 2}.
\end{eqnarray}
Then, we have
\begin{eqnarray}
{\cal A}_L&=&i\frac{G_F}{\sqrt 2}V_{CKM}f_2m_B^2A_0^{BV_1}(0),\\
{\cal A}_\parallel&=&-iG_FV_{CKM}f_2m_2 \left(m_B+m_1\right) A_1^{BV_1}(0), \\
{\cal A}_\perp&=&iG_FV_{CKM}f_2m_2\left(m_B-m_1\right)V^{BV_1}(0).
\end{eqnarray}
Any of the two vector mesons in the final sates will decay via strong interaction. They form two decay planes with various decay angles. Thus, for any given $B\to VV$ decay, one can define five typical observables corresponding to the branching fraction, two out of the three polarization fractions $f_L$, $f_\parallel, f_\perp$, and two relative phases $\phi_\parallel$, $\phi_\perp$:
\begin{equation} \label{eq:polobs}
  f^{ B}_{L,\parallel,\perp} = \frac{|{\cal A}_{L,\parallel,\perp}|^2}
  {  |{\cal A}_L|^2+|{\cal A}_\parallel|^2 + |{\cal A}_\perp|^2},
  \qquad\qquad
  \phi^{B}_{\parallel,\perp} = Arg\frac{{\cal A}_{\parallel,\perp}}{{\cal A}_0}.
\end{equation}
Apparently in the naive factorization, $f_L$ is expected to be close to unity from  eq.(\ref{herachy}).

CP symmetry is violated in weak interactions, thus one  expects to have different numbers for observables of   $B\to VV$ decay with those of its CP-conjugate $\overline B$ decay.    The CP averaged decay rate and the   CP asymmetry are then defined as
\begin{eqnarray}
   && \Gamma \equiv\frac{|\mathbf{p}|}{8\pi m_B^2} \frac{|\overline{\cal A}_L|^2+|\overline{\cal A}_\parallel|^2 + |\overline{\cal A}_\perp|^2 +|{\cal A}_L|^2+|{\cal A}_\parallel|^2 + |{\cal A}_\perp|^2}{2},\\
    &&A_{CP}\equiv \frac{|\overline{\cal A}_L|^2+|\overline{\cal A}_\parallel|^2 +|\overline{\cal A}_\perp|^2 -|{\cal A}_L|^2-|{\cal A}_\parallel|^2 - |{\cal A}_\perp|^2}{|\overline{\cal A}_L|^2+|\overline{\cal A}_\parallel|^2 + |\overline{\cal A}_\perp|^2 +|{\cal A}_L|^2+|{\cal A}_\parallel|^2 + |{\cal A}_\perp|^2},
\end{eqnarray}
where $\mathbf{p}$ is the 3-momentum of the final state vector meson in the rest frame of the $B$ meson.
Observables $f_h^{\overline B}$ and $\phi^{\overline B}_h$ are also defined as in (\ref{eq:polobs}), and CP averages and asymmetries are calculated by
\begin{equation}
    f_h \equiv \frac{1}{2} \left( f_h^{\overline B}+f_h^{B} \right),
    \qquad
    A_{CP}^{h} \equiv \frac{f_h^{\overline B}-f_h^{B}}{f_h^{\overline B}+f_h^{B}}
\end{equation}
($h=L,\parallel,\perp$) for the polarization fractions and
\begin{equation}
  \begin{aligned}
    \phi_h \equiv\frac{1}{2} \left( \phi_h^{\overline B}+\phi_h^{B} \right),  \qquad
    \Delta\phi_h \equiv \frac{1}{2} \left( \phi_h^{\overline B}-\phi_h^{B} \right),
  \end{aligned}\label{phasedef}
\end{equation}
for the phase observables $\phi_h$ and $\Delta\phi_h$. Unless otherwise indicated, for each observable quoted we imply the average of a process and its CP-conjugate one.

For the neutral $B$ meson decays, if $f=\bar f$,  the time-dependent of $CP$ violation is defined through
\begin{align}
{A}_{cp}^h(t)=\frac{\Gamma^h(\overline B^0(t)\to f)-\Gamma^h(B^0(t)\to f)}{\Gamma^h(\overline B^0(t)\to f)+\Gamma^h(B^0(t)\to f)}=S_{f}^h\mathrm {sin}(\Delta m_{B} t)-C_{f}^h\mathrm {cos}(\Delta m_{B} t),
\end{align}
where $\Delta m_{B}>0$ is the mass difference of the two neutral $B$ meson mass eigenstates. The definitions of two parameters $C_{f}^h$  and
$S_{f}^h$ are given as:
\begin{align}
C_{f}^h =\frac{1-|\lambda_f^h|^2}{1+|\lambda_f^h|^2}, \,\,\,\,\,\,
S_{f}^h =\frac{2Im(\lambda_f^h)}{1+|\lambda_f^h|^2} ,
\end{align}
where the parameters standing for the mixing of neural mesons read
\begin{align}
\lambda_f^h=\frac{q}{p}\frac{\bar{A}_f^h}{A_f^h}, \,\,\,\,\,\,
\frac{q}{p}=\frac{V^*_{tb}V_{tD}}{V_{tb}V^*_{tD}}(D=d,s) ,
\end{align}
$A_f^h$ is the decay amplitude of $B^{0}\to f^h$ and $\bar{A}_f^h$ is the the CP-conjugate one. Obviously, the $CP$ violations in $B\to VV$ are much complicated than those of $B\to PP$ and $B\to PV$ modes.

\begin{figure}
  \begin{center}
  \scalebox{0.4}{\epsfig{file=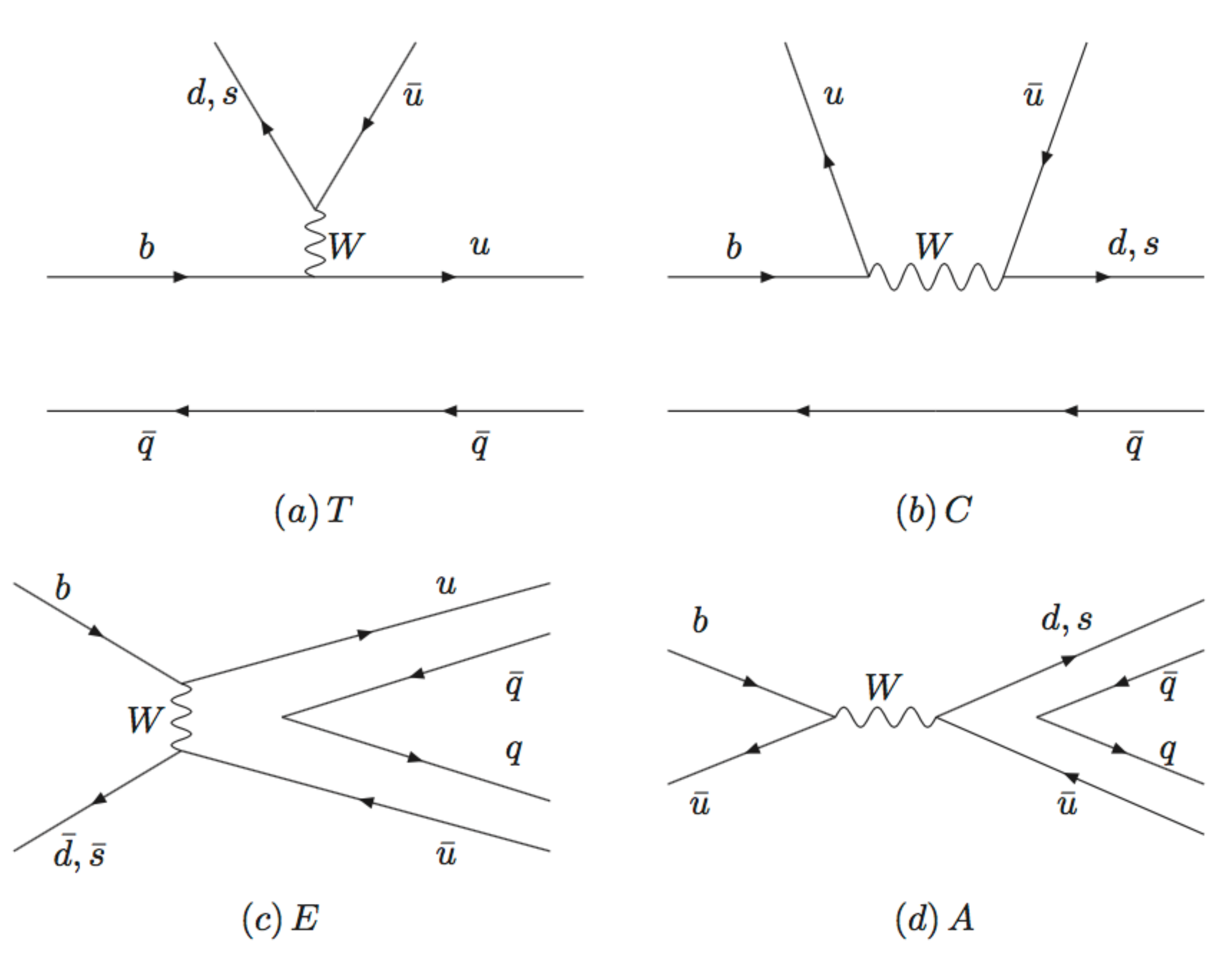}}
  \caption{Topological tree diagrams contributing to
     charmless $B\to VV$ decays:
  (a) the color-favored   tree emission diagram, $T$;
  (b) the color-suppressed  tree emission diagram, $C$;
  (c) the $W$-exchange     diagram, $E$ and
  (d) the $W$-annihilation     diagram, $A$.}
  \label{Treediagram}
  \end{center}
  \end{figure}

Now, we shall introduce the FAT approach. All of the hadronic B decays are weak decays, which are perturbatively calculable. However, all the initial and final states are hadrons, which involve non-perturbative QCD effects. The factorization of perturbative and non-perturbative QCD is limited to certain power expansion of $1/m_b$, which give limited accuracy of theoretical precision. The topological diagram approach does not rely on the QCD factorization, but group all the decay amplitudes by different Feynman diagrams according weak interaction. That means only  factorization of weak interaction from strong interaction is  required.  The QCD corrections to each weak decay diagram are extracted from experimental data, including all perturbative or non-perturbative ones. In this case, this approach is a kind of model independent method to deal with hadronic B decays.  Among these weak Feynman diagrams, we have      four kinds belong to the process induced by tree  diagram, which should be the leading contribution shown in Fig.\ref{Treediagram}.   They are denoted as
\begin{itemize}
  \item $T$, denoting the color-favored tree diagram with external W emission;
  \item $C$, denoting the color-suppressed tree diagram with internal W emission;
  \item $E$, denoting the W-exchange diagram;
  \item $A$, denoting the annihilation diagram.
 \end{itemize}

 In the conventional diagrammatic approach, the numerical results of each topology diagram can be fitted directly from the experimental data, by using the SU(3) symmetry. It is well known  that the breaking of SU(3) symmetry can reach $20-30\%$, which indicates that the prediction power of conventional diagrammatic approach is limited.
  For the decay processes dominated by the $T$-type diagram, such as $B \to D\pi$, the amplitudes can be expressed by the products of transition form factor, decay constant of the emitted meson and the short-distance dynamics Wilson coefficients \cite{Zhou:2015jba}, which has been proved in QCDF, PQCD, and SCET.  Results of all the three approaches agree with each other and with the experimental data well. Similarly, in our $B_{(s)} \to VV$ decays, the $T$ amplitudes of all three polarizations are expressed by
\begin{eqnarray}\label{Tree}
&&{T}^0=i\frac{G_F}{\sqrt 2}V_{CKM}a_1(\mu)f_2m_B^2A_0^{BV_1}(0),\\
&&{T}^\parallel=-iG_FV_{CKM}a_1(\mu)f_2m_2 \left(m_B+m_1\right) A_1^{BV_1}(0),  \\
&&{T}^\perp=iG_FV_{CKM}a_1(\mu)f_2m_2\left(m_B-m_1\right)V^{BV_1}(0).
\end{eqnarray}
In contrast to the conventional diagrammatic approach, where decay amplitude needs to be fitted from experiments, it is obvious that no free parameter is introduced in the $T$ amplitude.  $a_1(\mu)=C_2(\mu)+C_1(\mu)/3$ is the combination of the effective Wilson coefficients of four-quark operators \cite{Buchalla:1996vs},
  where $\mu=m_b /2=2.1 \mathrm{GeV}$ is the factorization scale. At this scale, $a_1(2.1 \mathrm{GeV})=1.05$.
    The SU(3) breaking effects are automatically kept, by different   decay constants and  form factors for different processes.

For the color-suppressed tree diagram shown in Fig.\ref{Treediagram}(b), the nonfactorizable contribution   is dominant, thus no factorization formula is given. The decay amplitude and strong phase need to be fitted from experimental data. However, we can still factorize out the  corresponding decay constant and form factor to keep  the SU(3) breaking effects  as an approximation:
\begin{eqnarray}
&&{ C}^0=i\frac{G_F}{\sqrt 2}V_{CKM}\chi_C^0 e^{i \phi^0_C}f_2m_B^2A_0^{BV_1}(0) \\
&&{ C}^\parallel=-iG_FV_{CKM}\chi_C^\parallel e^{i \phi^\parallel_C}f_2m_2 \left(m_B+m_1\right) A_1^{BV_1}(0) \\
&&{ C}^\perp=iG_FV_{CKM}\chi_C^\perp e^{i \phi^\perp_C}f_2m_2\left(m_B-m_1\right)V^{BV_1}(0).
\end{eqnarray}

Similarly, for the   $W$-exchange diagram shown in Fig.\ref{Treediagram}(c), we factorize out the meson decay constants to characterize  the SU(3) breaking effects. The  decay amplitudes for the three polarizations  are then written as:
\begin{eqnarray}
&&{E}^0=i\frac{G_F}{\sqrt 2}V_{CKM}f_Bf_1f_2m_B^2\chi_E^0e^{i \phi^0_E}\frac{1}{f_\rho^2},\\
&&{E}^\parallel=-iG_FV_{CKM}f_Bf_1f_2m_B^2\chi_E^\parallel e^{i \phi^\parallel_E}\frac{1}{f_\rho^2}, \\
&&{ E}^\perp=iG_FV_{CKM}f_Bf_1f_2m_B^2\chi_E^\perp e^{i \phi^\perp_E}\frac{1}{f_\rho^2},
\end{eqnarray}
where the amplitude $\chi_E$ is dimensionless with normalization to $f_\rho^2$.  As for the W-annihilation diagram $A$ shown in Fig.\ref{Treediagram}(d), its contribution is too small to be considered in the present work.
 In practice, even if we keep this contribution in our fitting program,  we cannot get a stable solution for them, since the  experimental precision is not good enough.

\begin{figure}
  \begin{center}
 \scalebox{0.4}{\epsfig{file=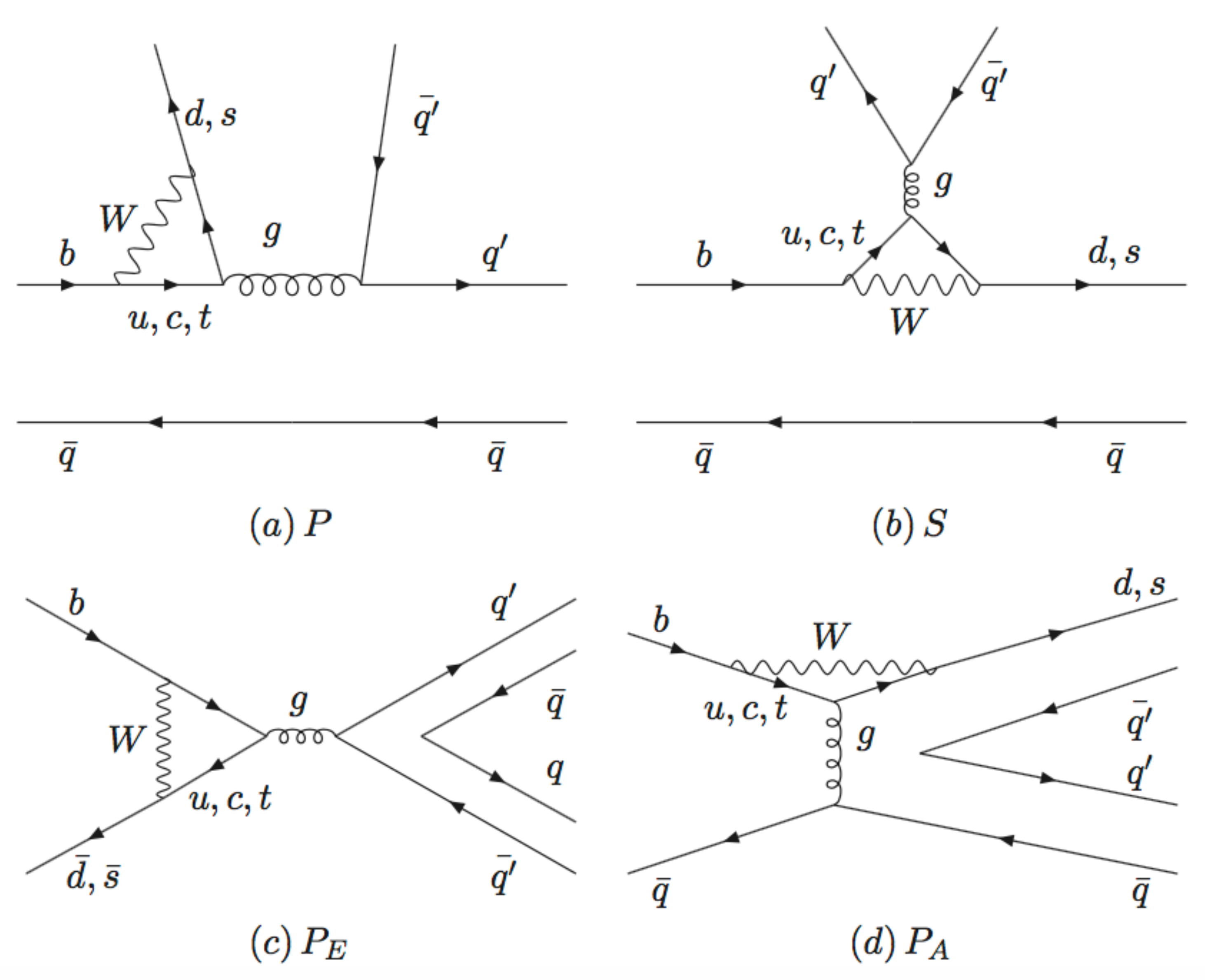}}
  \caption{Topological penguin diagrams contributing to
     charmless  $B\to VV$ decays:
  (a) the color-favored  QCD-penguin diagram, $P$;
  (b) the flavor-singlet   QCD-penguin diagram, $S$;
  (c) the time-like  penguin annihilation diagram, $P_E$ and
  (d) the    space-like penguin annihilation diagram, $P_A$.}
 \label{Penguindiagram}
  \end{center}
  \end{figure}

There are also   QCD-penguin  diagrams shown in Fig.\ref{Penguindiagram}. Although they are loop diagrams, which should be suppressed comparing with the tree level diagrams, they are enhanced by the large CKM matrix elements and large top quark mass. There are also four types of them:
 \begin{itemize}
    \item $P$, denoting the QCD penguin diagram;
  \item $S$, denoting the flavor-singlet QCD-penguin diagram;
  \item $P_E$, denoting the time-like penguin annihilation diagram;
  \item $P_A$, denoting the  space-like penguin annihilation diagram.
\end{itemize}

Similar to the color favored tree diagram $T$,  the QCD penguin diagram $P$ has the same $V-A$ structure, which can be proved factorization for all orders of $\alpha_s$ expansion. Therefore, we do not need to introduce free parameter for them. The decay amplitudes are again expressed by the products of transition form factor, decay constant of the emitted meson and the short-distance  Wilson coefficients:
 \begin{eqnarray}
&&{ P}^0=i\frac{G_F}{\sqrt 2}V_{CKM}a_4(\mu)f_2m_B^2A_0^{BV_1}(0), \\
&&{ P}^\parallel=-iG_FV_{CKM}a_4(\mu)f_2m_2 \left(m_B+m_1\right) A_1^{BV_1}(0,) \\
&&{ P}^\perp=iG_FV_{CKM}a_4(\mu)f_2m_2\left(m_B-m_1\right)V^{BV_1}(0).
\end{eqnarray}

Similar to the color suppressed tree diagram $C$, the flavor-singlet QCD penguin diagram $S$ are non-factorizable that expressed  by
\begin{eqnarray}
&&{ S}^0=i\frac{G_F}{\sqrt 2}V_{CKM}\chi_{S}^0e^{i \phi^0_{S}}f_2m_B^2A_0^{BV_1}(0), \\
&&{ S}^\parallel=-iG_FV_{CKM}\chi_{S}^\parallel e^{i \phi^\parallel_{S}}f_2m_2 \left(m_B+m_1\right) A_1^{BV_1}(0),  \\
&&{ S}^\perp=iG_FV_{CKM}\chi_{S}^\perp e^{i \phi^\perp_{S}}f_2m_2\left(m_B-m_1\right)V^{BV_1}(0).
\end{eqnarray}

Similar to the W-annihilation type diagram, the contribution of time-like penguin diagram shown in Fig.\ref{Penguindiagram}(c) is also negligible, which can not be fitted from the experimental data at the current precision. Lastly, the penguin  annihilation diagram (space-like) contributions $P_A$ are   expressed as:
\begin{eqnarray}
&&P_A^0=i\frac{G_F}{\sqrt 2}V_{CKM}f_Bf_1f_2m_B^2\chi_{P_A}^0e^{i \phi^0_{P_A}}\frac{1}{f_\rho^2},\\
&&P_A^\parallel=-iG_FV_{CKM}f_Bf_1f_2m_B^2\chi_{P_A}^\parallel e^{i \phi^\parallel_{P_A}}\frac{1}{f_\rho^2},\\
&&P_A^\perp=iG_FV_{CKM}f_Bf_1f_2m_B^2\chi_{P_A}^\perp e^{i \phi^\perp_{P_A}}\frac{1}{f_\rho^2}.\label{PAE}
\end{eqnarray}

Now, we have 12   magnitudes and 12  phases to be fitted simultaneously from the the  experimental data. With the fitted  parameters, we can predict all other  $B_q \to VV$ ($q=u,d,s$) decays with  branching fractions, polarization fractions, and CP asymmetry paramters.

\section{$\chi^2$ Fitting and Numerical Results}\label{sec-3}

To characterize the flavor SU(3) breaking effects in our calculation, we need    input parameters  of various meson masses \cite{Agashe:2014kda} and decay constants that are summarized in   Table \ref{mess and dc}. Unlike the meson masses, the value of decay constants is not known in experiment, but can be given from theoretical calculation, such as QCD sum rules \cite{Baker:2013mwa}, Bethe-Salpeter equation \cite{Maris:1999nt}, and lattice QCD \cite{Gray:2005ad}, and we taken the value from \cite{Zhou:2016jkv} with $5\%$ uncertainty.

For the calculation of color favored tree diagram $T$ and QCD penguin digram $P$, we also need the input  of form factors for $B \to V$ transitions.   There are many calculations for form factors, such as light-cone sum rules \cite{sumrule}, perturbative QCD approach \cite{pqcd}, lattice QCD \cite{lqcd} etc. The central values we used in this work of the transition form factors at $q^2=0$ are shown in Table \ref{form factors}. To estimate the theoretical uncertainty of the numerical results in our calculation, we include the uncertainties of all form factors as large as $10\%$. In fact, they are one of the major sources of theoretical uncertainty in our numerical results, especially for processes dominated by the color favored tree ($T$) diagram.  Since the  final state meson mass is small comparing with the large B meson mass, the $q^2$ dependence of form factors will be neglected. In fact, the effects of $q^2$ dependence to numerical results are negligible, which has been indicated in $B\to PP,PV$ decays \cite{Zhou:2016jkv}.

For the CKM matrix elements, we adopt the Wolfenstein parametrization, which are given as \cite{Agashe:2014kda}
\begin{eqnarray}
\lambda = 0.22537 \pm 0.00061,\,\,\, A = 0.814^{+0.023}_{-0.024},\,\,\,
\bar \rho = 0.117 \pm 0.021,\,\,\, \bar \eta = 0.353\pm 0.013.
\end{eqnarray}
 The life time of $B$ mesons are taken from the particle data group \cite{Agashe:2014kda}, given as
\begin{eqnarray}
\tau_{B^\pm}=1.641 ps,\,\,\,\tau_{B^0}=1.519 ps,\,\,\,\tau_{B_s^0}=1.497 ps.
\end{eqnarray}

\begin{table}
\caption{The mass and decay constant of meson (in units of GeV)}
\centering
\begin{tabular}{ccc}
\hline
\hline
Meson & ~~~~~~Mass~~~~~~ & ~~~~~~Decay Constant~~~~~~\\
\hline
$B^{\pm/0}$   &5.28 & 0.190\\
$B_s^{0}$     &5.36 & 0.225\\
$\rho$        &0.77 & 0.213\\
$\omega$      &0.78 & 0.192\\
$\phi$        &1.01 & 0.225\\
$K^*$         &0.89 & 0.220\\
\hline
\hline
\end{tabular}\label{mess and dc}
\end{table}

\begin{table}
\caption{The transition form factor of $B\to V$ at $q^2=0$}
\centering
\begin{tabular}{lccccc}
\hline\hline
&$B \to \rho$ &$B \to K^*$ & $B\to\omega$ & $B_s\to K^*$ &$B_s\to\phi$ \\
\hline
$V(0)$  &0.33&0.41&0.29&0.31&0.42\\
$A_0(0)$&0.32&0.38&0.28&0.36&0.44\\
$A_1(0)$&0.25&0.29&0.22&0.23&0.31\\
\hline\hline
\end{tabular}\label{form factors}
\end{table}

\subsection{The $\chi^2$ fit for theoretical parameters}

Unlike $B\to D^{(*)}M$ and $B\to PP(V)$ decay modes, the amplitudes of $B\to VV$ modes are much complicated, because each decay process has three polarization contributions, which means that the number of parameters will increase threefold. From previous section, we do not introduce any new parameter in color favored tree diagram $T$ and QCD penguin diagram $P$. For the color-suppressed tree diagram $C$,  the flavor-singlet QCD-penguin diagram $S$,  the $W$-exchange diagram and QCD-penguin  annihilation diagram $P_A$,  we have all together 24 parameters $\chi_C^{0,\parallel,\perp}$, $\phi_C^{0,\parallel,\perp}$, $\chi_{S}^{0,\parallel,\perp}$, $\phi_{S}^{0,\parallel,\perp}$, $\chi_{E}^{0,\parallel,\perp}$,  $\phi_{E}^{0,\parallel,\perp}$,       $\chi_{P_A}^{0,\parallel,\perp}$ and $\phi_{P_A}^{0,\parallel,\perp}$. So many free parameters are difficult to be determined from the limited number of experimental measurements. It will also decrease the prediction power of this FAT approach. As indicated from the QCD factorization approach \cite{Cheng:2009cn,Cheng:2009mu} and the perturbative QCD approach \cite{Zou:2015iwa} calculations, the color suppressed tree diagram $C$, $W$ exchange diagram $E$ and the flavor singlet QCD penguin diagram $S$ are dominated by longitudinal polarization contributions.  For simplicity, we here drop the negligible transverse contributions  of these topological diagrams to set
$\chi_C^{\parallel}=\chi_C^{\perp}=0$,
$\chi_E^{\parallel}=\chi_E^{\perp}=0$, 
and $\chi_{S}^{\parallel}=\chi_{S}^{\perp}=0$. 
Therefore, the only transverse polarization amplitudes to be fitted are from the penguin annihilation diagram. According to the power counting of this diagram \cite{Kagan:2004uw}, the negative helicity amplitude is Chirally enhanced while the positive helicity amplitude is still suppressed (to be neglected). This results in a relation of  $\chi_{P_A}^{\parallel} \approx \chi_{P_A}^{\perp}$.
Thus, there are only 10 universal parameters left, which will be fitted by experimental data.

In the experimental sides, after the first decay modes $B\to \phi K^*$ were measured, more and more observables have been measured, involving the branching fractions, CP asymmetries, polarization fractions, and relative phases of helicity amplitudes. Since some observables are measured with very poor precision, the data with less than $3\sigma$ significance will not be used in our fitting program. Then, we have 46 experimental data, involving 18 branching fractions, 20 polarization fractions, 6 relative phases, and 2 direct CP asymmetries.

In the $\chi^2$ fitting, in term of $46$ experimental observables $x_i\pm\Delta x_i$ and the corresponding theoretical predictions $x_i^{\rm th}$, the $\chi^2$ function can be  defined as
\begin{align}
\chi^{2}=\sum_{i=1}^{46}\left(\frac{x_i^{\rm th}-x_i}{\Delta x_i}\right)^2 .
\end{align}
With the amplitudes and data,  we can extract the 10 parameters by minimizing the $\chi^2$.
The best-fitted values of the parameters are given as:
\begin{eqnarray}
   &\chi_C^{0}=0.23\pm 0.05,\,\,\,\,\, \phi_C^{0}=0.48\pm0.29;\,\,\,\,\, \chi_E^{0}=0.082\pm0.026,\,\,\,\,\, \phi_E^{0}=1.69\pm0.16;\nonumber\\
   &\chi_{S}^{0}=0.018\pm 0.003,\,\,\,\,\, \phi_{S}^{0}=1.29\pm0.22;\,\,\,\,\, \chi_{P_A}^{0}=0.012\pm0.002,\,\,\,\,\, \phi_{P_A}^{0}=-0.07\pm0.18;\nonumber\\
&\chi_{P_A}^{\parallel,\perp}=0.0098\pm 0.0003,\,\,\,\,\, \phi_{P_A}^{\parallel,\perp}=-0.21\pm0.09;\label{fitresults}
\end{eqnarray}
The $\chi^2/\mathrm{d.o.f}=82.0/(46-10)$ is $2.28$. Compared with the corresponding fitted values of $B\to PP$ and $PV$ decays \cite{Zhou:2016jkv}, the magnitudes of the longitudinal polarization $\chi_i^0(i=C,E,S,P_A)$ are at the similar size.

\subsection{Numerical Results and Discussions}

For the convenience of later discussion, we shall class 33 decay modes into 5 categories. The first category is $T$-dominated decays, involving four decay channels, $\overline B^0 \to \rho^+\rho^-$, $B^- \to \rho^+\rho^0$, $B^- \to \rho^+\omega$ and $\overline B_s^0 \to \rho^-K^{*+}$. Five decay process $\overline B^0 \to \rho^0\rho^0$, $\overline B^0 \to \rho^0\omega$,  $\overline B_s^0 \to \rho^0K^{*0}$ and $\overline B_s^0 \to \omega K^{*0}$ dominated by $C$ diagram,   fall into the second class. The third class is dominated by QCD penguin diagram $P$, with eleven decays $B\to\rho K^*$, $\omega K^*$, $\phi K^*$, $K^*K^*$ and $B_s\to K^* K^*$, $\phi K^*$ and $\phi\phi$. There are three more decay channels $B^-\to K^{*0} K^{*-}$, $\overline B^0\to K^{*0}\overline K^{*0}$   and $\overline B_s^0\to \phi K^{*0}$, which are also $P$ dominated. But their branching ratios are highly suppressed comparing with the previous 11 channels, since they are mediated by $b\to d$ transition instead of $b\to s$ transition, which are CKM suppressed.
The fourth class includes $B^-\to \rho^-\phi$, $B^0\to \rho^0\phi$, $B^0\to \omega\phi$, $B_s\to \rho^0\phi$ and $B_s\to \omega\phi$, to which the flavor-singlet QCD-penguin mainly contribute. The last group is five  decay channels dominated by annihilation type diagrams: $B^0 \to K^{*+}K^{*-}$, $B_s \to \rho^{+}\rho^{-}$, $B_s \to \rho^{0}\rho^{0}$, $B_s \to \rho^{0}\omega$ and  $B_s \to \omega\omega$.

\begin{table}
\begin{center}
\caption{Branching fractions and the direct $CP$ asymmetries of  $\overline B \to VV$ decay modes.  The  experimental data \cite{Agashe:2014kda} are also given for comparison. }\label{Tab:br}
\renewcommand*{\arraystretch}{0.8}
\begin{tabular}{c|l| l |l|l| l }
\hline
\hline
 \multirow{2}{*}{Class} & \multirow{2}{*}{Decay Mode} &
 \multicolumn{2}{c|}{Branching Fraction / $10^{-6}$} &
 \multicolumn{2}{c}{$A_{CP}$ / percent} \\
 \cline{3-6}
&  & Theory  & Exp & Theory    & Exp \\
 \hline
 \multirow{4}{*}{T}& $B^-\to\rho^-\rho^0$                & $21.7\pm1.8\pm4.2\pm2.2$    & $24.0\pm 1.9$  & $0$ & $-5\pm5$ \\
&$\overline B^0\to\rho^-\rho^+$      & $29.5\pm1.9\pm5.4\pm3.0$    & $28.3\pm 2.1$  & $-8.10\pm2.94$   \\
&$B^-\to\rho^-\omega$                & $18.2\pm1.5\pm2.8\pm1.6$    & $15.9\pm 2.1$  & $-3.45\pm5.38$ & $-20\pm9$ \\
&$\overline B_s^0\to K^{*+}\rho^-$  & $38.6\pm0.1\pm7.3\pm3.9$ & $ $       & $-10.9\pm3.0$ & $ $ \\
\hline
 \multirow{5}{*}{C}&$\overline B^0\to\rho^0\rho^0$      & $0.94\pm0.46\pm0.11\pm0.14$ & $0.97\pm 0.24$ & $49.7\pm13.4$   \\
&$\overline B^0\to\rho^0\omega$      & $1.48\pm0.71\pm0.06\pm0.20$ & $<1.6$         & $-38.5\pm13.6$  \\
&$\overline B^0\to\omega\omega$      & $1.20\pm0.49\pm0.12\pm0.18$ & $1.2\pm0.4$    & $28.1\pm13.8$ & $ $ \\
&$\overline B_s^0\to K^{*0}\rho^0$  & $1.18\pm0.39\pm0.21\pm0.12$ & $<767$ & $4.9\pm18.3$ & $ $ \\
&$\overline B_s^0\to K^{*0}\omega$  & $0.97\pm0.33\pm0.16\pm0.10$ & $ $    & $32.2\pm16.0$ & $ $ \\
\hline
 \multirow{11}{*}{P}&$B^-\to\rho^-\overline K^{*0}$              & $10.4\pm1.6\pm1.7\pm1.1$ & $9.2\pm1.5$  & $1.00\pm0.17$ & $-1\pm16$ \\
&$B^-\to\rho^0 K^{*-}$                       & $5.83\pm0.66\pm0.76\pm0.65$ & $4.6\pm1.1$  & $34.6\pm8.3$       &$31\pm13$ \\
&$\overline B^0\to\rho^0\overline K^{*0}$    & $5.09\pm0.75\pm0.82\pm0.53$     & $3.9\pm1.3$  & $-0.6\pm4.0$ & $-6\pm9$ \\
&$\overline B^0\to\rho^+ K^{*-}$             & $10.5\pm1.3\pm1.4\pm1.2$ & $10.3\pm2.6$ & $34.3\pm6.3$ & $21\pm15$ \\
&$B^-\to\omega K^{*-}$                       & $4.24\pm0.70\pm0.32\pm0.51$ & $<7.4$       & $-30.1\pm13.8$ & $29\pm35$ \\
&$\overline B^0\to\omega\overline K^{*0}$    & $3.10\pm0.75\pm0.27\pm0.38$ & $2.0\pm0.5$  & $-11.7\pm4.0$ & $45\pm25$ \\
&$B^-\to \phi K^{*-}$                        & $9.31\pm1.90\pm1.83\pm0.97$ & $10\pm2$ & $1.00\pm0.27$ & $-1\pm8$ \\
&$\overline B^0\to \phi\overline K^{*0}$     & $8.64\pm1.76\pm1.70\pm0.90$ & $10\pm0.5$ & $1.00\pm0.27$ & $0\pm4$ \\
&$\overline B_s^0\to \overline K^{*0} K^{*0}$  & $14.9\pm2.0\pm1.9\pm2.3$ & $28\pm7$ & $0.78\pm0.19$ & $ $ \\
&$\overline B_s^0\to   K^{*-} K^{*+}$         & $15.9\pm1.7\pm1.7\pm2.6$ & $ $              & $21.1\pm7.1$ & $ $ \\
&$\overline B_s^0\to \phi \phi$    & $26.4\pm4.8\pm4.5\pm3.8$    & $19.3\pm3.1$  & $0.83\pm0.28$ & $ $ \\
 \hline
 \multirow{3}{*}{P}&$B^-\to K^{*0} K^{*-}$                      & $0.66\pm0.10\pm0.13\pm0.08$ & $1.2\pm0.5$    & $-24.8\pm2.6$ & $ $ \\
&$\overline B^0\to K^{*0}\overline K^{*0}$   & $0.61\pm0.09\pm0.12\pm0.07 $ &               & $-24.8\pm2.6$ & $ $ \\
&$\overline B_s^0\to \phi K^{*0}$  & $0.70\pm0.11\pm0.13\pm0.08$ & $1.13\pm0.30$ & $-17.3\pm5.6$ & $ $ \\
\hline
 \multirow{5}{*}{$S$}&$B^-\to \phi \rho^{-}$              & $0.06\pm0.02\pm0.01\pm0.01$ & $<3.0$  & $0$ & $ $ \\
&$\overline B^0\to \phi\rho^{0}$     & $0.03\pm0.01\pm0.01\pm0.00$ & $<0.33$ & $0$ & $ $ \\
&$\overline B^0\to \phi\omega$       & $0.02\pm0.01\pm0.00\pm0.002$ & $<0.7$  & $0$ & $ $ \\
&$\overline B_s^0\to \phi \rho^0$  & $0.07\pm0.03\pm0.01\pm0.01$ & $<617$ & $0$ & $ $ \\
&$\overline B_s^0\to \phi \omega$  & $3.69\pm1.19\pm0.74\pm0.37$ &        & $-15.0\pm7.0$ & $ $ \\
\hline
 \multirow{5}{*}{E($P_E$)}&$\overline B^0\to K^{*+} K^{*-}$            & $1.43\pm0.91\pm0\pm0.29$    & $<2.0$         & $0$ & $ $ \\
&$\overline B_s^0\to\rho^-\rho^+$  & $0.10\pm0.06\pm0\pm0.02$ & $ $    & $0$ & $ $ \\
&$\overline B_s^0\to\rho^0\rho^0$  & $0.05\pm0.03\pm0\pm0.01$ & $<320$ & $0$ & $ $ \\
&$\overline B_s^0\to\rho^0\omega$  & $0.08\pm0.05\pm0\pm0.01$ & $ $ & $0$ & $ $ \\
&$\overline B_s^0\to\omega\omega$  & $0.03\pm0.02\pm0\pm0.01$ & $ $ & $0$ & $ $ \\
\hline
\hline
\end{tabular}
\end{center}
\end{table}

  In Table~\ref{Tab:br}, we list the branching fractions, and the theoretical errors correspond to the uncertainties due to variation of (i) the fitted universal $\chi$ values, (ii) the heavy-to-light form factors and (iii) the uncertainty of decay constants. The error of the variation of the CKM matrix elements is negligible. We note that for other observable, we have combined these uncertainties by adding them in quadrature and show the resulting uncertainty, due to to the space limitations in the tables.
 The decay $B \to \phi\phi$ is absent  in our tables, because it is a pure  annihilation type process induced only by time-like penguin diagram $P_E$, which contribution is neglected in this work, though its branching fraction is estimated to be at the order of $10^{-8}$ based on PQCD approach \cite{liying:phihi}. Comparing our predictions with experimental data, one can find that our results can accommodate the data well, within uncertainties from both theoretical and experimental sides. For those decays that have not been measured, our prediction can be tested in the ongoing LHCb experiment and the forthcoming Belle-II experiment.

For the color-suppressed decay mode $\overline B^0\to \rho^0\omega$, its branching fraction is predicted to be $0.08\times 10^{-6}$ and  $0.4\times 10^{-6}$ in QCDF \cite{Cheng:2008gxa} and in PQCD \cite{Zou:2015iwa}, respectively. However, in this work, we predicted its branching ratio to be  $1.48\times 10^{-6}$, which is about 18 times larger than the result of QCDF. Moreover, in both QCDF and PQCD, the longitudinal fraction is calculated to be about $67\%$, which is   smaller than our prediction ($87\%$). The measurement of this mode in future will help us to understand the dynamics of this decay mode. For $\overline B\to \omega\omega$, although the experimental data agree quite well with our value, but there exist larger uncertainties in both experimental and theoretical results. So, the precise measurement of $B\to \omega\omega$ is necessary, too.

\begin{table}[tbh!]
\begin{center}
\caption{The polarization fractions and  relative phases of $\overline B \to VV$ decay modes.  The  experimental data are taken from \cite{Agashe:2014kda}. }\label{Tab:pf}
\renewcommand*{\arraystretch}{0.8}
{\footnotesize
\begin{tabular}{l| l |l|l| l|l| l|l| l }
 \hline
 \hline
 \multirow{2}{*}{Decay Mode} &
 \multicolumn{2}{c|}{$f_L$/ percent} &
 \multicolumn{2}{c|}{$f_\perp$/ percent}&
 \multicolumn{2}{c|}{$\phi_\parallel$/ rad}&
 \multicolumn{2}{c}{$\phi_\perp$/ rad} \\
 \cline{2-9}
  & Theory  & Exp & Theory    & Exp& Theory    & Exp& Theory    & Exp \\
 \hline
$B^-\to\rho^-\rho^0$                & $95.5\pm1.1$ & $95\pm1.6$      & $2.22\pm0.64$ &      &$-0.09\pm0.05$ &       & $-0.09\pm0.05$  \\
$\overline B^0\to\rho^-\rho^+$      & $92.6\pm1.6$ & $98.8\pm2.6$    & $3.65\pm0.91$ &      & $-0.27\pm0.08$ &     & $-0.27\pm0.08$  \\
$B^-\to\rho^-\omega$                & $92.7\pm1.4$ & $90\pm6$         & $3.60\pm0.76$ &      & $-0.23\pm0.07$ &    & $-0.23\pm0.07$ \\
$\overline B_s^0\to K^{*+}\rho^-$  & $94.4\pm1.2$ &   & $2.74\pm0.64$ &  & $ -0.08\pm0.03$    &   & $-0.08\pm0.03$\\
\hline
$\overline B^0\to\rho^0\rho^0$      & $81.7\pm10.8$ & $60\pm23$        & $9.21\pm5.50$ &      & $-0.04\pm0.44$ &     & $-0.03\pm0.44$ \\
$\overline B^0\to\rho^0\omega$      & $82.7\pm9.5$ &                  & $ 8.68\pm4.82$ &     & $0.98\pm0.22$ &     & $0.98\pm0.22$ \\
$\overline B^0\to\omega\omega$      & $92.2\pm3.6$ &                  & $3.94\pm1.85$ &      & $-1.46\pm0.26$ &     & $-1.45\pm0.26$ \\
$\overline B_s^0\to K^{*0}\rho^0$  & $79.8\pm8.0$ &    & $10.2\pm4.1$ &   & $-0.94\pm0.28$    &   & $-0.94\pm0.28$\\
$\overline B_s^0\to K^{*0}\omega$  & $77.9\pm9.2$ &    & $11.2\pm4.7$ &   & $-0.73\pm0.31$    &   & $-0.73\pm0.31$\\
\hline
$B^-\to\rho^-\overline K^{*0}$              & $46.0\pm12.9$ & $48\pm8$    & $27.2\pm7.0$ &      & $2.07\pm0.22$ &      & $2.08\pm0.22$\\
$B^-\to\rho^0 K^{*-}$                       & $40.7\pm10.6$ & $78\pm12$    & $29.8\pm5.9$ &      & $2.24\pm0.20$ &     & $2.24\pm0.20$\\
$\overline B^0\to\rho^0\overline K^{*0}$    & $48.7\pm12.3$ & $40\pm14$   & $25.8\pm6.7$ &      & $2.08\pm0.21$ &      & $2.09\pm0.21$\\
$\overline B^0\to\rho^+ K^{*-}$             & $38.9\pm11.3$ & $38\pm13$   & $30.8\pm6.3$ &      & $2.18\pm0.22$ &     & $2.18\pm0.22$\\
$B^-\to\omega K^{*-}$                       & $29.9\pm6.8$ & $41\pm19$   & $35.3\pm4.5$ &      & $0.02\pm0.85$ &      & $0.03\pm0.85$\\
$\overline B^0\to\omega\overline K^{*0}$    & $29.4\pm17.5$ & $69\pm13$   & $35.6\pm9.4$ &     & $-2.62\pm0.53$ &      & $-2.61\pm0.53$
\\
$B^-\to \phi K^{*-}$                     & $48.0\pm16.0$ & $50\pm5$     & $25.9\pm8.6$ & $20\pm5$& $2.47\pm0.27$ & $2.34\pm0.18$ & $2.47\pm0.27$ &$2.58\pm0.17$\\
$\overline B^0\to \phi\overline K^{*0}$  & $48.0\pm16.0$ & $49.7\pm1.7$ & $26.0\pm8.6$ & $22.4\pm1.5$ & $2.47\pm0.27$ & $2.43\pm0.11$ & $2.47\pm0.27 $ &$2.53\pm0.09$\\
$\overline B_s^0\to \overline K^{*0} K^{*0}$ & $34.3\pm12.6$ &    & $33.2\pm6.9$  & $38\pm11$  & $2.10\pm0.23$ &  & $2.10\pm0.23$\\
$\overline B_s^0\to   K^{*-} K^{*+}$         & $30.9\pm10.4$  &    & $34.9\pm5.8$ &            & $2.19\pm0.22$&  & $2.20\pm0.22$\\
$\overline B_s^0\to \phi \phi$    & $39.7\pm16.0$  & $36.2\pm1.4$ & $31.2\pm8.9$ & $30.9\pm1.5$  &$2.53\pm0.28$ &$2.55\pm0.11$ &$2.56\pm0.27$ &$2.67\pm0.23$\\
 \hline
$B^-\to K^{*0} K^{*-}$                        & $58.3\pm11.1$ & $75\pm21$   & $20.8\pm6.0$ &     & $2.10\pm0.20$   &     & $2.09\pm0.20$\\
$\overline B^0\to K^{*0}\overline K^{*0}$     & $58.3\pm11.1$ & $80\pm13$   & $20.8\pm6.0$ &     & $2.10\pm0.20$    &     & $2.09\pm0.20$\\
$\overline B_s^0\to \phi K^{*0}$  & $38.9\pm14.7$  & $51\pm17$    & $31.4\pm8.1$ &               &$2.52\pm0.27$  &              & $2.55\pm0.27$& \\
\hline
\hline
\end{tabular}
}
\end{center}
\end{table}


  The polarization fractions, as well as relative phases, are shown in Table~\ref{Tab:pf}.
One can find that for the first four tree-dominated processes, they fully respect the helicity hierarchy that are dominated by longitudinal polarization. The major theoretical uncertainties here are from the heavy-light form factors and the CKM matrix element $|V_{ub}|$. On the contrary, for the color suppressed tree diagram (C) dominant decays, their branching fractions become much smaller due to the absence of $T$-type diagrams. Correspondingly, the largest uncertainties are from the $\chi_C$, in particular from the transverse polarizations, which have also been confirmed in QCD factorization \cite{Beneke:2006hg, Cheng:2008gxa}. For the three $B\to \rho\rho$ decay modes, the decay amplitudes can be read:
\begin{eqnarray}
&& \sqrt{2} A(B^-\to\rho^- \rho^0) =T+C;\\
&&A(\overline B^0\to\rho^+ \rho^-)=T+E+P+P_A;\\
&&A(\overline B^0\to\rho^0 \rho^0)=-C+E+P+P_A.
\end{eqnarray}
The three decay amplitudes make an isospin triangle. For illustration, we list the numerical  results of longitudinal polarization for each topological diagram of these decays,
\begin{eqnarray}
|T^{B\to\rho_L\rho_L}|:|C^{B\to\rho_L\rho_L}|:|E^{B\to\rho_L\rho_L}|:|P^{B\to\rho_L\rho_L}|:|P_A^{B\to\rho_L\rho_L}|=1:0.22:0.21:0.14:0.08.
\end{eqnarray}
For the decay  $B^-\to\rho^- \rho^0$, although the absolute value of the $C$ diagram is suppressed,  it can enhance the  magnitude of the decay amplitude by $20\%$.   With the larger contribution from $C$ diagram, the large branching fraction of $\overline B^0\to\rho^0 \rho^0$ is also explained. By including the transverse momenta of inner quarks, the $\overline B^0\to\rho^0 \rho^0$ has been calculated in perturbative QCD approach in ref.\cite{Zou:2015iwa}, where they got a rather smaller branching fractions ($0.27\times 10^{-6}$) and a smaller longitudinal fraction ($12\%$). In fact, there is a large discrepacy between experimental measurements from BaBar and Belle, and the number we quoted in this work is the naive averaged value. So, it is very important to have a refined measurement of the branching fractions and  longitudinal fractions of $B\to\rho\rho$ decays to draw the final conclusion.

For $B_s$ decays dominated by $T$ and $C$ diagrams, they have the same manner as $B$ decays, for example, the color-allowed decay $\overline B_s^0 \to \rho^-K^{*+}$ has large branching fraction  and large  longitudinal fraction, while the branching fraction of color suppressed $\overline B_s^0 \to K^{*0}\rho^0(\omega)$ decays is a bit smaller and the transverse polarizations are about $20\%$. Comparing the branching fractions of $\overline B_s^0 \to \rho^-K^{*+}$ with $\overline B^0\to\rho^-\rho^+$, we find that the former is larger than the latter one. The reason is that the form factor $A_0^{B_s \to K^*}$ is larger than $A_0^{B \to \rho}$ by $13\%$. Considering the life time difference, the large gap between these two branching fractions can be well understood. Since the order of these three branching fractions is about $10^{-6}$,     they should be measurable in the running LHCb experiment.

We now  discuss the 14 decay modes dominated by QCD penguin diagrams $P$. Among these decays, 11 decays induced by $b\to s$ transition have branching fractions up to $10^{-5}$ due to large CKM matrix elements $|V_{tb}V_{ts}^*|$.   Some of them have been measured precisely in the experiments, including branching fractions, polarization fractions,  and even CP asymmetries. The  first measured one, and also the most well measured channel is  $B\to \phi K^*$ modes that are induced by $b\to ss\bar s$ transition.  In this decay, the magnitudes of QCD penguin diagram $P$, the flavor-singlet penguin diagram $S$ and the penguin annihilation diagram $P_A$ are at the same order magnitude.  For illustration,  numerical results of each diagram are provided in Table~\ref{VVam}.  It is easy to see that   the penguin annihilation diagram has a very large transverse polarization contribution.  In QCDF \cite{Beneke:2006hg, Cheng:2008gxa}, in order to increase the effects of annihilation diagrams,  the free parameters    $\rho_A$ and $\phi_A$ introduced for the power suppressed penguin annihilation diagram are required to be very large. In the so-called PQCD approach, the large effects of annihilation are arrived, by including the transverse momenta of inner quarks. So, we then conclude that the larger transverse polarizations in $B\to \phi K^*$ arise from the annihilation diagrams.

\begin{table}
\begin{center}
\caption{Amplitudes of each diagram of $B\to \phi K^*$ ($\times 10^{-8} \mathrm{GeV}$)}\label{VVam}
\renewcommand*{\arraystretch}{0.8}
\begin{tabular}{c ccc}
 \hline
   &$P$ &$S$ & $P_A$    \\
\hline
$A^0$
& $-1.36+5.01 i$
& $1.39-0.43 i$
& $-0.21-2.29 i$   \\
$A^\parallel$
& $-0.33+1.22 i$
& $0+0 i$
& $-0.60-2.60 i$   \\
$A^\perp$
& $-0.33+1.23 i$
& $0+0 i$
& $-0.60 -2.60 i$   \\
\hline
\end{tabular}
\end{center}
\end{table}

Another decay mode induced by $b\to ss\bar s$ is $B_s \to \phi\phi$. From Tables.~\ref{Tab:br} and \ref{Tab:pf}, one can see that although our estimation of branching fractions agree with data within uncertainties, our center value is a bit larger than the experimental data, and the predicted polarization fractions are in agreement with the data. The acceptable divergency in branching fraction is due to the larger form factor $A_0^{B_s \to \phi}$ we adopted, which in fact is also related to the branching ratio of $B_s \to \phi K^*$.  If we consider the $B_s \to \phi \phi$ alone,  the present data favor a smaller $A_0^{B_s \to \phi}$. In this point, the precisely calculations of form factor in lattice QCD and other effective approach are needed.

For $B^- \to \rho^-\phi$, $\overline B^0 \to \rho^0\phi$ and $\overline B^0 \to \omega^0\phi$, they are dominated by the flavor-singlet QCD penguin diagram. These tree decay modes have an identical amplitude $S$. Compared with the longitudinal amplitudes,  both transverse amplitudes are so small that can be neglected safely, so that the longitudinal polarizations are about $100\%$. It should be emphasized that the neglected electroweak penguin will enhance the negative helicity amplitudes, and the longitudinal polarization fraction will decrease to $70\%$ \cite{Beneke:2005we}. For decays $\overline B_s^0 \to \rho^0\phi$ and $\overline B_s^0 \to \omega^0\phi$, they have large branching fractions as they are governed by the larger CKM elements. In fact, because we drop the electroweak contribution away, and 
the flavor singlet penguins is cancelled,  the $\overline B_s^0 \to \rho^0\phi$ decay is a mode with only  color-suppressed tree diagram (C) contribution. While for the $\overline B_s^0 \to \omega^0\phi$, both $C$ and $S$ contribute, so its branching fraction is larger. Honestly, for this decays, there is another larger uncertainty we have not included here. In this work, we have assumed the ideal mixing, which means that there is no $s\bar s$ component in the $\omega$ component. As we mentioned above, $B_s\to \phi\phi$ has a large branching fraction, even a small mixing angle will affect the predictions remarkably \cite{Li:piphi}.

\begin{table}
\begin{center}
\caption{The predictions of the $CP$ asymmetries of longitudinal and perpendicular polarizations, as well as relative phase differences of $B\to VV$ decays.}\label{Tab:ob}
\renewcommand*{\arraystretch}{0.8}
\begin{tabular}{l| l |l|l | l }
 \hline
 \hline
 Decay Mode &
$A_{CP}^0$/ percent &
$A_{CP}^\perp$/ percent&
$\Delta\phi_\parallel$/ rad&
$\Delta\phi_\perp$/ rad \\
 \hline
 $B^-\to\rho^-\rho^0$                & $0$            &  $0$          &  $0$              & $0$         \\
$\overline B^0\to\rho^-\rho^+$      & $1.30\pm0.54$  & $-16.3\pm8.2$ &  $-0.41\pm0.05$   &  $-0.41\pm0.05$  \\
$B^-\to\rho^-\omega$                & $2.38\pm0.86$      & $-30.2\pm11.6$       & $-0.70\pm0.07$   & $-0.72\pm0.07$ \\
$\overline B_s^0\to K^{*+}\rho^-$  & $0.91\pm0.45$  & $-15.4\pm9.5$   & $-0.52\pm0.06$  & $-0.54\pm0.06$\\
\hline
$\overline B^0\to\rho^0\rho^0$      & $10.5\pm9.6$       & $-46.9\pm13.9$     & $1.89\pm0.19$     & $1.89\pm0.19$ \\
$\overline B^0\to\rho^0\omega$      & $-8.68\pm7.72$      & $41.6\pm13.3$      & $-1.47\pm0.09$    & $-1.47\pm0.09$ \\
$\overline B^0\to\omega\omega$      & $2.10\pm1.87$     & $-24.7\pm14.1$     & $-1.43\pm0.07$     & $-1.43\pm0.07$ \\
$\overline B_s^0\to K^{*0}\rho^0$  & $0.47\pm4.69$   & $-1.89\pm18.3$          & $2.03\pm0.10$  & $2.02\pm0.10$\\
$\overline B_s^0\to K^{*0}\omega$  & $8.37\pm7.28$   & $-29.5\pm16.3$   & $-1.58\pm0.13$& $-1.58\pm0.13$\\
\hline
$B^-\to\rho^-\overline K^{*0}$              & $1.40\pm0.56$     & $ -1.19\pm0.19$   & $-0.01\pm0.00$    & $-0.01\pm0.00$\\
$B^-\to\rho^0 K^{*-}$                       & $35.0\pm19.8$      & $-24.2\pm9.0$     & $-0.82\pm0.15$     & $-0.82\pm0.15$\\
$\overline B^0\to\rho^0\overline K^{*0}$    & $-0.41\pm4.3$     & $0.39\pm4.06$    & $0.13\pm0.04$     & $0.13\pm0.04$\\
$\overline B^0\to\rho^+ K^{*-}$             & $37.2\pm18.9$     & $-23.8\pm6.9$     & $-0.66\pm0.12$     & $-0.65\pm0.12$\\
$B^-\to\omega K^{*-}$                       & $-93.4\pm25.9$     & $39.6\pm13.0$    & $2.09\pm0.90$    & $2.09\pm0.90$\\
$\overline B^0\to\omega\overline K^{*0}$    & $-27.7\pm19.4$     & $11.5\pm4.0$     & $-0.04\pm0.12$    & $-0.04\pm0.12$\\
$B^-\to \phi K^{*-}$                     & $1.26\pm0.71$        & $-1.16\pm0.30$ &  $-0.02\pm0.00$     &$-0.02\pm0.00$  \\
$\overline B^0\to \phi\overline K^{*0}$  & $1.26\pm0.71$        & $-1.16\pm0.30$ &  $-0.02\pm0.00$ & $-0.02\pm0.00$\\
$\overline B_s^0\to \overline K^{*0} K^{*0}$  & $1.81\pm0.69$  & $-0.94\pm0.20$  & $-0.005\pm0.003$  & $-0.004\pm0.003$\\
$\overline B_s^0\to   K^{*-} K^{*+}$          & $32.6\pm20.2$  & $-14.8\pm7.3$  & $-0.70\pm0.14$     & $-0.70\pm0.14$\\
$\overline B_s^0\to \phi \phi$   & $1.55\pm0.85$  & $-1.02\pm0.29$ & $-0.01\pm0.00$ & $-0.01\pm0.00$\\
 \hline
$B^-\to K^{*0} K^{*-}$                     & $-20.2\pm8.0$     & $28.3\pm3.3$     & $0.16\pm0.04$     &  $0.16\pm0.04$\\
$\overline B^0\to K^{*0}\overline K^{*0}$  & $-20.2\pm8.0$     & $28.3\pm3.3$     & $0.16\pm0.04$     & $0.16\pm0.04$\\
$\overline B_s^0\to \phi K^{*0}$  & $-32.9\pm15.0$ & $21.0\pm5.7$ & $0.27\pm0.08$& $0.27\pm0.08$\\
\hline
\hline
\end{tabular}
\end{center}
\end{table}

In Table \ref{Tab:pf}, as we expected, the longitudinal polarization fractions of the  tree diagram dominant  decays are predicted to be near unity with errors in the $(5-10)\%$ range. The CP asymmetries in the longitudinal polarizations of these decays are less than $5\%$, as shown in Table~\ref{Tab:ob}. Although the CP asymmetries of the perpendicular polarizations of these decays shown in Table~\ref{Tab:ob} are large, they are difficult to measure, since their fractions are too small.  For the decays controlled by the $C$ diagram, although the longitudinal polarization factions become smaller, they still play the primary roles with large uncertainties. Furthermore, the uncertainties of $A^0_{CP}$ is large, though some of them can reach $10\%$. Our  theoretical result of $f_L$  for the $B^-\to \rho^-\rho^0$  is a bit larger than the data, but the situation of $\overline B ^0 \to \rho^+\rho^-$ is in the opposite direction.

Because each $B \to VV$ decay  has three polarizations, the possible time-dependent $CP$ violation is complicated,  it is very hard to measure them precisely in the experiments. For the Tree-dominant decays, the transverse parts can be neglected, and the measurement of time-dependence $CP$ violation of these decays becomes plausible. In this work, we calculated the $C_{f}$ and $S_{f}$ of the longitudinal parts of $\overline B^0 \to \rho^+\rho^-$ and $\overline B^0 \to \rho^0\rho^0$, as shown in Table~\ref{Tab:cpv}. Obviously, our results agree with experiment, though there are large uncertainties in both theoretical and experimental sides. Also, we present the numerical results of $ \bar B^{0}_s \to (K^{\ast -} K^{\ast +})_L$ and $ \bar B^{0}_s \to(\phi \phi)_L$, which may be measured in future, as both modes have large branching fractions and the final states are easy to be identified. Noted that the precise measurement of $C_{\rho\rho}^L$ and $S_{\rho\rho}^L$ will help us to determine the CKM angles $\alpha$ and $\gamma$ \cite{Beneke:2006rb}.

\begin{table}
\begin{center}
\caption{Prediction of the time-dependent CP Violation($\%$). }\label{Tab:cpv}
\begin{tabular}{lcccc}
 \hline
 \multirow{2}{*}{Decay Mode} &
 \multicolumn{2}{c }{$S_f$ } &
 \multicolumn{2}{c }{$C_f$ }
 \\
 \cline{2-5}
  & Theory  & Exp & Theory    & Exp \\
 \hline
$ \bar B^{0} \to (\rho^{-} \rho^{+})_L$ & $-3.67\pm3.02$& $-6\pm17$  & $6.80\pm3.12$ &$-5\pm13$\\
$ \bar B^{0} \to (\rho^{0} \rho^{0})_L$ & $41.7\pm27.8$ & $30\pm70$ & $-57.2\pm17.4$ &$20\pm90$\\
$ \bar B^{0} \to(\omega \omega)_L$ & $15.7\pm13.3$ & & $-30.0\pm15.1$ &\\
$ \bar B^{0}_s \to (K^{\ast -} K^{\ast +})_L$ & $80.9\pm14.4$ &  &$-50.2\pm20.7$ &\\
$ \bar B^{0}_s \to(\phi \phi)_L$ & $2.16\pm0.76$ &  &$-2.39\pm0.82$ &\\
\hline
\end{tabular}
\end{center}
\end{table}

We then come to decay modes $B \to \rho K^*$ and $B_s \to K^* K^*$. For the pure-penguin process $B^- \to \rho^- \overline K^{*0}$,  which is similar to the decays $B \to \phi K^*$,  the  penguin annihilation diagram will give large transverse polarization fraction. For the $B^- \to \rho^0 K^{*-}$, to which the tree operators also contribute, the destructive interference between tree and penguin operators reduce the longitudinal amplitude. So, the smaller longitudinal polarization fraction of $B^- \to \rho^0 K^{*-}$ is obtained. The large longitudinal polarization fraction $f_L$ of this decay   is only measured by BABAR experiment. We hope Belle or Belle II experiment can help to resolve this puzzle. Due to the factor $1/\sqrt 2$, the branching fraction of $B^- \to \rho^0 K^{*-}$ is about half of that of $B^- \to \rho^- \overline K^{*0}$. The analysis and the result of the modes $B \to \omega K^*$ and $B_s \to K^* K^*$ should be similar to those of $B \to \rho K^*$. From Tables.~\ref{Tab:pf} and \ref{Tab:ob}, we find that for all penguin dominant decays $f_\parallel\approx f_\perp$, $\phi_\parallel \approx \phi_\perp$ and $\Delta\phi_\parallel \approx \Delta\phi_\perp$, which indicates that the positive-helicity amplitudes are about zero. In fact, due to the suppression of leading QCD penguins, $\rho K^*$ final states have also been used to prob electroweak penguin effect \cite{Beneke:2006hg}, however, this kind of contributions have been neglected in the present work due to not enough experimental data.

As for the last five pure annihilation type decays, they all have two kinds of contributions: the W exchange diagram (E) and the time-like penguin annihilation diagram ($P_E$). As discussed in previous section, there are not enough experimental data  to determine the amplitude of time-like penguin annihilation diagram ($P_E$). In our fitting, we have to set it to 0. Since all the $B_s$ decays in this category are dominated by this $P_E$ contribution except $\overline B_s^0 \to \rho^0\omega$, due to the small CKM matrix elements in W exchange diagram (E), the branching fractions of these decay modes    are not stable in Table~\ref{Tab:br} with large uncertainties.   Only $\overline B^0\to K^{*+} K^{*-}$ decay has a relatively larger CKM matrix elements in the W exchange diagram (E), which makes a relatively larger branching fractions, but still with large uncertainty. And the pure W exchange diagram (E) channel $\overline B_s^0 \to \rho^0\omega$ also have large uncertainty because of the large error of $\chi_E^0$. Therefore, we conclude that    the current experimental data can not help us to make predictions on this kind of decays, but waiting for   the running LHCb experiment, Belle-II or other future colliders.

\section{Summary}\label{sec-4}

In the work, we preformed analysis of 33 charmless two-body $B_{(s)} \to VV$ decays  within the factorization-assisted topological-amplitude approach. In contrast to the charmless $B\to PP$ and $B\to PV$ decays, more parameters (triple number in principle) are needed to describe the three polarization amplitudes of $B_{(s)} \to VV$ decays. However, with the current   46 experimental data, we can only fit 10 universal  parameters of them. For the decays with large transverse polarization fractions, such as the   penguin diagram contribution dominated decays, we need only one transverse polarization amplitude in penguin annihilation diagram to explain all the polarization data.   We calculated many decay modes not yet measured, involving the branching fractions, the polarization fractions, $CP$ violation parameters, as well as    relative strong phases. These results will be tested in the LHCb experiment and future Belle-II experiment.

\section*{Acknowledgement}

We are grateful to  W. Wang, F.-S. Yu,  S-H Zhou, Y.-B. Wei, J.-B. Liu, X.-D.  Gao and Q. Qin for helpful discussions. The work is partly supported by National Natural Science Foundation of China (11575151, 11375208, 11521505, 11621131001 and 11235005) and the Program for New Century Excellent Talents in University (NCET) by Ministry of Education of P. R. China (NCET-13-0991). Y. Li and C.D. L\"u are also supported by the Open Project Program of State Key Laboratory of Theoretical Physics, Institute of Theoretical Physics, Chinese Academy of Sciences, China (No.Y5KF111CJ1).  Y.Li is also support
by the Natural Science Foundation of Shandong Province (Grant No. ZR2016JL001).



\begin{thebibliography}{99}
\bibitem{Bevan:2014iga}
  A.~J.~Bevan {\it et al.} [BaBar and Belle Collaborations],
  Eur.\ Phys.\ J.\ C {\bf 74}, 3026 (2014)
  [arXiv:1406.6311 [hep-ex]].

\bibitem{Buchalla:2008jp}
  M.~Artuso {\it et al.},
  Eur.\ Phys.\ J.\ C {\bf 57}, 309 (2008)
  [arXiv:0801.1833 [hep-ph]].

\bibitem{Aushev:2010bq}
  T.~Aushev {\it et al.},
  arXiv:1002.5012 [hep-ex].

\bibitem{Browder:2007gg}
  T.~Browder, M.~Ciuchini, T.~Gershon, M.~Hazumi, T.~Hurth, Y.~Okada and A.~Stocchi,
  JHEP {\bf 0802}, 110 (2008)
  [arXiv:0710.3799 [hep-ph]].

\bibitem{Gianotti:2002xx}
  F.~Gianotti {\it et al.},
  Eur.\ Phys.\ J.\ C {\bf 39}, 293 (2005)
  [hep-ph/0204087].

\bibitem{LHCb:2011dta}
  The LHCb Collaboration [LHCb Collaboration],
  CERN-LHCC-2011-001.
\bibitem{cepc}  { http://cepc.ihep.ac.cn}

\bibitem{Wirbel:1985ji}
  M. Bauer, B. Stech, M. Wirbel, Z. Phys. C34 (1987) 103;\\
  Ahmed Ali, G. Kramer, Cai-Dian Lu,  Phys. Rev. D58 (1998) 094009;\\
  Ahmed Ali, G. Kramer, Cai-Dian Lu,  Phys. Rev. D59 (1999) 014005.

\bibitem{Beneke:2000ry}
 M.~Beneke, G.~Buchalla, M.~Neubert and C.~T. Sachrajda, Nucl. Phys. {\bf B591}, 313 (2000), [hep-ph/0006124];\\
 M.~Beneke and M.~Neubert, Nucl. Phys. {\bf B675}, 333 (2003), [hep-ph/0308039].

\bibitem{Lu:2000em}
  C.~D.~Lu, K.~Ukai and M.~Z.~Yang,   Phys.\ Rev.\ D {\bf 63}, 074009 (2001);\\
  Y.~Y.~Keum, H.~N.~Li and A.~I.~Sanda, Phys.\ Rev.\ D {\bf 63}, 054008 (2001)

\bibitem{Bauer:2000yr}
  C.~W.~Bauer, S.~Fleming, D.~Pirjol and I.~W.~Stewart,  Phys.\ Rev.\ D {\bf 63}, 114020 (2001)  [hep-ph/0011336];\\
  C.~W.~Bauer, D.~Pirjol, I.~Z.~Rothstein and I.~W.~Stewart, Phys.\ Rev.\ D {\bf 70}, 054015 (2004)

\bibitem{diagramapp}
L. L. Chau and H. Y. Cheng, Phys. Rev. Lett. 56, 1655 (1986);
L. L. Chau and H. Y. Cheng, Phys.\ Rev.\ D {\bf 36}, 137 (1987);
L. L. Chau, H. Y. Cheng, W. K. Sze, H. Yao and B. Tseng,Phys.\ Rev.\ D {\bf 43}, 2176 (1991).

\bibitem{cwchiangB}
M. Gronau, O. F. Hernandez, D. London and J. L. Rosner, Phys.\ Rev.\ D {\bf 50}, 4529 (1994);
M. Gronau, O. F. Hernandez, D. London and J. L. Rosner, Phys. Phys.\ Rev.\ D {\bf 52}, 6374 (1995);
C. W. Chiang, M. Gronau, Z. Luo, J. L. Rosner and D. A. Suprun, Phys.\ Rev.\ D {\bf 69}, 034001 (2004)
C. W. Chiang, M. Gronau, J. L. Rosner and D. A. Suprun, Phys.\ Rev.\ D {\bf 70}, 034020 (2004) ;
C. W. Chiang and Y. F. Zhou, JHEP 0612, 027 (2006) [hep-ph/0609128];
C. W. Chiang and Y. F. Zhou, JHEP 0903, 055 (2009) [arXiv:0809.0841 [hep-ph]].
H. Y. Cheng, C. W. Chiang and A. L. Kuo,  Phys.\ Rev.\ D {\bf 91},014011 (2015)

\bibitem{Cheng:2012xb}
H.~Y.~Cheng and C.~W.~Chiang,  Phys.\ Rev.\ D {\bf 86}, 014014 (2012)  [arXiv:1205.0580 [hep-ph]].
H.~Y.~Cheng, C.~W.~Chiang and A.~L.~Kuo,  Phys.\ Rev.\ D {\bf 93}, no. 11, 114010 (2016)  [arXiv:1604.03761 [hep-ph]].

\bibitem{Li:2012cfa}
H.~N.~Li, C.~D.~L\"u and F.~S.~Yu,  Phys.\ Rev.\ D {\bf 86}, 036012 (2012)   [arXiv:1203.3120 [hep-ph]].
H.~N.~Li, C.~D.~L\"u, Q.~Qin and F.~S.~Yu,  Phys.\ Rev.\ D {\bf 89}, no. 5, 054006 (2014)  [arXiv:1305.7021 [hep-ph]].

\bibitem{Zhou:2015jba}
S.~H.~Zhou, Y.~B.~Wei, Q.~Qin, Y.~Li, F.~S.~Yu and C.~D.~Lu,  Phys.\ Rev.\ D {\bf 92}, no. 9, 094016 (2015)  [arXiv:1509.04060 [hep-ph]].

\bibitem{Zhou:2016jkv}
S.~H.~Zhou, Q.~A.~Zhang, W.~R.~Lyu and C.~D.~L¨¹,  arXiv:1608.02819 [hep-ph].


\bibitem{Koerner1979}
J.~G. K{\"o}rner and G.~R. Goldstein,
Phys. Lett. {\bf B79}, 105 (1979).

\bibitem{Li:2003he}
X.~Q. Li, G.-r. Lu and Y.~D. Yang,
 Phys. Rev. {\bf D68}, 114015 (2003), [hep-ph/0309136].

\bibitem{Kagan:2004uw}
A.~L. Kagan,
 Phys. Lett. {\bf B601}, 151 (2004), [hep-ph/0405134].

\bibitem{Beneke:2006hg}
  M.~Beneke, J.~Rohrer and D.~Yang,
  Nucl.\ Phys.\ B {\bf 774}, 64 (2007)
  [hep-ph/0612290].

\bibitem{Bartsch:2008ps}
  M.~Bartsch, G.~Buchalla and C.~Kraus,
  arXiv:0810.0249 [hep-ph].

\bibitem{Cheng:2008gxa}
  H.~Y.~Cheng and K.~C.~Yang,
  Phys.\ Rev.\ D {\bf 78}, 094001 (2008)
  [arXiv:0805.0329 [hep-ph]].

\bibitem{Cheng:2009cn}
  H.~Y.~Cheng and C.~K.~Chua,
  Phys.\ Rev.\ D {\bf 80}, 114008 (2009)
  [arXiv:0909.5229 [hep-ph]].

\bibitem{Cheng:2009mu}
  H.~Y.~Cheng and C.~K.~Chua,
  Phys.\ Rev.\ D {\bf 80}, 114026 (2009)
  [arXiv:0910.5237 [hep-ph]].

\bibitem{Li:2004ti}
  H.~n.~Li and S.~Mishima,
  Phys.\ Rev.\ D {\bf 71}, 054025 (2005)
  [hep-ph/0411146].

\bibitem{Ali:2007ff}
  A.~Ali, G.~Kramer, Y.~Li, C.~D.~Lu, Y.~L.~Shen, W.~Wang and Y.~M.~Wang,
  Phys.\ Rev.\ D {\bf 76}, 074018 (2007)
  [hep-ph/0703162 [HEP-PH]].

\bibitem{Zou:2015iwa}
  Z.~T.~Zou, A.~Ali, C.~D.~Lu, X.~Liu and Y.~Li,
  Phys.\ Rev.\ D {\bf 91}, 054033 (2015)
  [arXiv:1501.00784 [hep-ph]].

\bibitem{Hou:2004vj}
W.-S. Hou and M.~Nagashima,
hep-ph/0408007.

\bibitem{Ladisa:2004bp}
M.~Ladisa, V.~Laporta, G.~Nardulli and P.~Santorelli,
Phys. Rev. {\bf D70}, 114025 (2004), [hep-ph/0409286].

\bibitem{Yang:2004pm}
Y.-D. Yang, R.-M. Wang and G.-R. Lu,
Phys. Rev. {\bf D72}, 015009 (2005), [hep-ph/0411211].

\bibitem{Das:2004hq}
P.~K. Das and K.-C. Yang,
Phys. Rev. {\bf D71}, 094002 (2005), [hep-ph/0412313].

\bibitem{Kim:2004wq}
C.~S. Kim and Y.-D. Yang,
hep-ph/0412364.

\bibitem{Zou:2005gw}
W.-j. Zou and Z.-j. Xiao,
Phys. Rev. {\bf D72}, 094026 (2005), [hep-ph/0507122].

\bibitem{Huang:2005if}
H.-W. Huang {\em et~al.},
Phys. Rev. {\bf D73}, 014011 (2006), [hep-ph/0508080].

\bibitem{Baek:2005jk}
S.~Baek, A.~Datta, P.~Hamel, O.~F. Hernandez and D.~London,
Phys. Rev. {\bf D72}, 094008 (2005), [hep-ph/0508149].

\bibitem{Huang:2005qb}
C.-S. Huang, P.~Ko, X.-H. Wu and Y.-D. Yang,
Phys. Rev. {\bf D73}, 034026 (2006), [hep-ph/0511129].


\bibitem{Bao:2008hd}
  S.~S.~Bao, F.~Su, Y.~L.~Wu and C.~Zhuang,
  Phys.\ Rev.\ D {\bf 77}, 095004 (2008)
  [arXiv:0801.2596 [hep-ph]].


\bibitem{Buchalla:1996vs}
 G.~Buchalla, A.~J.~Buras, and M.~E.~Lautenbacher, Rev.\ Mod.\ Phys. {\bf 68} (1996) 1125 [hep-ph/9512380].

\bibitem{Li:2005kt}
  H.~n.~Li, S.~Mishima and A.~I.~Sanda,  Phys.\ Rev.\ D {\bf 72}, 114005 (2005)
  [hep-ph/0508041].

\bibitem{Agashe:2014kda}
  C.~Patrignani {\it et al.} [Particle Data Group],
  Chin.\ Phys.\ C {\bf 40} (2016) no.10,  100001.
  doi:10.1088/1674-1137/40/10/100001

\bibitem{liying:phihi}
  Y. Li, Phys.\ Rev.\ D {\bf 89}, 014013 (2014)

\bibitem{Beneke:2006rb}
  M.~Beneke, M.~Gronau, J.~Rohrer and M.~Spranger,
  Phys.\ Lett.\ B {\bf 638}, 68 (2006)
  [hep-ph/0604005].

  \bibitem{Beneke:2005we}
  M.~Beneke, J.~Rohrer and D.~Yang,
  Phys.\ Rev.\ Lett.\  {\bf 96}, 141801 (2006)
  [hep-ph/0512258]; \\
   C.-D. L\"u, Y.-L. Shen and W.~Wang,
   Chin. Phys. Lett. {\bf 23}, 2684 (2006).

  \bibitem{Li:piphi}
  Y.~Li, C.-D. L\"u, and W.~Wang,
  Phys.\ Rev.\ D {\bf 80}, 014024 (2009)

  \bibitem{Baker:2013mwa}
  Baker {\it et al.}, J. High Energy Phys. 07(2014)032 [arXiv:1310.0941[hep-ph]];
  P. Gelhausen {\it et al.}, Phys. Rev. D {\bf 88}, 014015 (2013) Erratum: [Phys. Rev. D {\bf 89}, 099901 (2014)] Erratum: [Phys. Rev. D {\bf 91}, 099901 (2015)] [[arXiv:1305.5432[hep-ph]]
  
  \bibitem{Maris:1999nt}
  P. Maris and P. C. Tandy, Phys. Rev. {\bf C60}, 055214 (1999) [nucl-th/9905056];
  Z. G. Wang {\it et al.}, Phys. Lett.  {\bf B584} 71 (2004). [hep-ph/0311150]  
  
  \bibitem{Gray:2005ad}
  G. Alan {\it et al.} [HPQCD Collaboration], Phys. Rev. Lett. {\bf 95}, 212001 (2005) [hep-lat/0507015];
  A. Bazavov {\it et al.} [Fermilab Lattice and MILC Collaborations], Phys. Rev. D {\bf 85}, 114506 (2012) [arXiv:1112.3051 [hep-lat]];
  H.Na {\it et al.}, Phys. Rev. D {\bf 86}, 034506 (2012) [arXiv:1202.4914]
  
  \bibitem{sumrule}
  P. Ball and R. Zwicky, Phys. Rev. {\bf D71}, 014029 (2005) [hep-ph/0412079];
  N. Khodjamirian, T. Mannel and N. Offen, Phys. Rev. {\bf D75}, 054013 (2007) [hep-ph/0611193];
  A. Bharucha, D. M. Straub and R. Zwicky, JHEP 1608 (2016) 098 [arXiv:1503.05534[hep-ph]]
  
  \bibitem{pqcd}
  T. Kurimoto, H. n. Li and and A. I. Sanda, Phys. Rev. {\bf D65}, 014007 (2002) [hep-ph/0105003];
  C. D. Lu and M. Z. Zhang, Eur. Phys. J. C {\bf 28}, 515 (2003) [hep-ph/0212373];
  R. H. Li and C. D. Lu, Phys. Rev. {\bf D79}, 034014 (2009) [arXiv:0901.0307[hep-ph]]
  
  \bibitem{lqcd}
  E. Dalgic {\it et al.}, Phys. Rev. {\bf D73}, 074502 (2006) Erratum: [Phys. Rev. {\bf D75}, 119906 (2007)] [hep-lat/0601021];
  R. R. Horgan, Z. Liu, S. Meinel and M. Wingate, Phys. Rev. {\bf D89}, 094501 (2014) [arXiv:1310.3722[hep-lat]];
  R. R. Horgan, Z. Liu, S. Meinel and M. Wingate, Pos LATTICE2014, 372 (2015) [arXiv:1501.00367[hep-lat]]
  
\end{thebibliography}
\end{document}